\documentclass[aps,prd,twocolumn,groupedaddress,amsmath,amssymb]{revtex4-1}
\usepackage{graphicx}  
\usepackage{dcolumn}   
\usepackage{bm}        
\usepackage{verbatim}   
\usepackage[utf8]{inputenc}
\usepackage{color}

\begin{document}

\title{Polarized Heavy Quarkonium Production in the Color Evaporation Model}
\author{Vincent Cheung}
\affiliation{
   Department of Physics,
   University of California, Davis,
   Davis, CA 95616, USA
   }
\author{Ramona Vogt}
\affiliation{
   Nuclear and Chemical Sciences Division,
   Lawrence Livermore National Laboratory,
   Livermore, CA 94551, USA
   }
\affiliation{
   Department of Physics,
   University of California, Davis,
   Davis, CA 95616, USA
   }
\date{\today}
\begin{abstract}
We explore polarized heavy quarkonium production using the color evaporation model at leading order. We present the polarized to total yield ratio as a function of center of mass energy and rapidity in $p+p$ collisions. At energies far above the $Q \overline Q$ production threshold, we find charmonium and bottomonium production to be longitudinally polarized ($J_z=0$).  The quarkonium states are also longitudinally polarized at central rapidity, becoming transversely polarized ($J_z=\pm1$) at the most forward rapidities.
\end{abstract}


\maketitle


\section{Introduction}
Even more than 40 years after the discovery of $J/\psi$, the production mechanism of quarkonium is still not well understood. Most recent studies of quarkonium production employ nonrelativistic QCD (NRQCD) \citep{CASWELL1986437}, which is based on an expansion of the cross section in the strong coupling constant and the $Q \overline Q$ velocity \citep{Bodwin:1994jh}.  The cross section is factorized into hard and soft contributions and divided into different color and spin states.  Each color state carries a weight, the long distance matrix elements (LDMEs) that are typically adjusted to the data above some minimum transverse momentum, $p_T$, value.  The NRQCD cross section has been calculated up to next-to-leading order (NLO).  The LDMEs, conjectured to be universal, fail to describe both the yields and polarization simultaneously for $p_T$ cuts less than twice the mass of the quarkonium state.  The polarization is sensitive to the $p_T$ cut:  the cut $p_T > 10$~GeV was chosen to describe both the yield and polarization in Ref.~\citep{Bodwin:2014gia} while $p_T > 3m$ was chosen for the excited states $\psi(2S)$ and $\Upsilon(3S)$ in Ref.~\citep{Faccioli:2014cqa} to fit the polarization.  The universality of the LDMEs can be tested by using those obtained at high $p_T$ to calculate the $p_T$-integrated cross section.  In Ref.~\citep{Feng:2015cba}, the $p_T$-integrated NRQCD cross section is calculated with LDMEs obtained with $p_T$ cuts in the range $3 < p_T < 10$ GeV.  The resulting midrapidity cross sections, $d\sigma/dy|_{y=0}$, systematically overshoot the $J/\psi$ data.  The lowest $p_T$ cut is most compatible with $d\sigma/dy|_{y=0}$ while calculations based on higher $p_T$ cuts can be up to an order of magnitude away from the data \citep{Feng:2015cba}. More recent analysis has shown that the $\eta_c$ $p_T$ distributions calculated with LDMEs obtained from $J/\psi$ yields using heavy quark spin symmetry \citep{HQSS1,HQSS2,HQSS3}, overshoots the high $p_T$ LHCb $\eta_c$ results \citep{Butenschoen:2014dra}.

The Color Evaporation Model (CEM) \citep{Barger:1979js,Barger:1980mg,Gavai:1994in,Ma:2016exq}, which considers all $Q \overline{Q}$ ($Q$ = $c$, $b$) production regardless of the quarks' color, spin, and momentum, is able to predict both the total yields and the rapidity distributions with only a single normalization parameter\citep{NVF}.  The CEM has so far only been used to predict spin-averaged quarkonium production: the polarization was not considered before. This paper presents a leading order (LO) calculation of quarkonium polarization in the CEM, a $p_T$-integrated result.  Currently, there are no exclusive NLO polarized $Q \overline Q$ calculations on which to impose the $H\overline{H}$ ($H$ = $D$, $B$) mass threshold. Our calculation is a first step toward a full CEM polarization result that provides a general idea of whether there is any appreciable LO polarization that might carry through to the next order even though the kinematics are different.  We will begin to address the $p_T$ dependence in a further publication.

In the CEM, all quarkonium states are treated the same as $Q\overline{Q}$ below the $H\overline{H}$ threshold where the invariant mass of the heavy quark pair is restricted to be less than twice the mass of the lowest mass meson that can be formed with the heavy quark as a constituent. The distributions for all quarkonium family members are assumed to be identical.  (See Ref.~\cite{Ma:2016exq} for a new treatment of the CEM $p_T$ distributions based on mass-dependent thresholds.)  In a $p+p$ collision, the production cross section for a quarkonium state is given by
\begin{eqnarray}
\label{cem_sigma}
\sigma &=& F_Q \sum_{i,j} \int^{4m_H^2}_{4m_Q^2}d\hat{s} \int dx_1 dx_2 f_{i/p}(x_1,\mu^2) f_{j/p}(x_2,\mu^2) \nonumber \\
&\times& \hat{\sigma}_{ij}(\hat{s}) \delta(\hat{s}-x_1 x_2 s) \;,
\end{eqnarray}
where $i$ and $j$ are $q$, $\overline q$ and $g$ such that $ij = q\overline{q}$ or $gg$.  The square of the heavy quark pair invariant mass is $\hat{s}$ while the square of the center-of-mass energy in the $p+p$ collision is $s$.  Here $f_{i/p}(x,\mu^2)$ is the parton distribution function (PDF) of the proton as a function of the fraction of momentum carried by the colliding parton $x$ at factorization scale $\mu$ and $\hat{\sigma}_{ij}$ is the parton-level cross section. Finally, $F_{Q}$ is a universal factor for the quarkonium state and is independent of the projectile, target, and energy. At leading order, the rapidity distribution, $d\sigma/dy$, is 
\begin{eqnarray}
\label{cem_rapdity}
\frac{d\sigma}{dy} &=& F_{Q} \int^{4m_H^2}_{4m_Q^2} \frac{d\hat{s}}{s} \Big\{ f_{g/p}(x_1,\mu^2) f_{g/p}(x_2,\mu^2) \hat{\sigma}_{gg} (\hat{s}) \nonumber \\
&+& \sum_{q=u,d,s} [f_{q/p}(x_1,\mu^2)f_{\overline{q}/p}(x_2,\mu^2) \nonumber \\ 
&+& f_{\overline{q}/p}(x_1,\mu^2)f_{q/p}(x_2,\mu^2)] \hat{\sigma}_{q\overline{q}} (\hat{s}) \Big\} \;,
\end{eqnarray}
where $x_{1,2} = (\sqrt{\hat{s}/s}) \exp(\pm y)$. We take the square of the factorization and renormalization scales to be $\mu^2 = \hat{s}$.


\section{Polarized $Q\overline{Q}$ production at the parton level}
\begin{figure}
\centering
\includegraphics[width=\columnwidth]{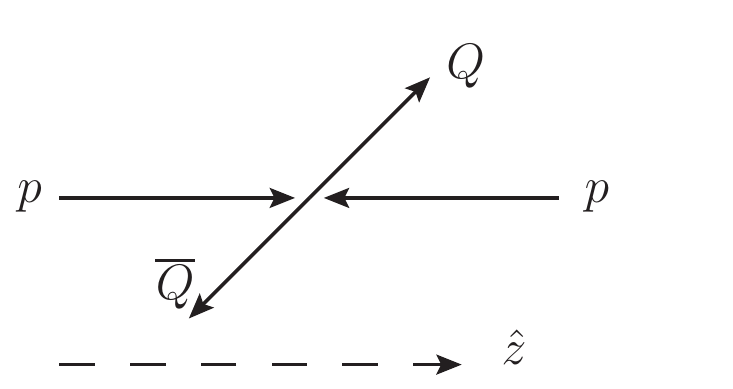}
\caption{\label{polarization} Orientation of z-axis indicated by the dashed arrowed line. Two proton arrows indicate the incoming beam directions. If the final state heavy quark-antiquark pair have the same helicity, then the total angular momentum along the $z$-axis, $J_z$, is 0 while if they have opposite helicity, then $J_z = \pm 1$.}
\end{figure}

At the parton level, the leading order calculation forces the final state $Q \overline{Q}$ pair to be produced back-to-back with zero total transverse momentum. We define the polarization of the $Q \overline{Q}$ pair to be either transversely polarized ($J_z =\pm 1$) or longitudinally polarized  ($J_z = 0$) in the helicity frame where the $z$ axis is pointing from $\overline{Q}$ to $Q$ along the beam axis as shown in Fig.~\ref{polarization}. Note that we are not distinguishing the $S=1$ triplet state from the $S=0$ singlet state. This will be addressed in a future publication, together with the separation into orbital angular momentum, $L$, states.

At leading order, there are four Feynman diagrams to consider, one for $q \overline q$ annihilation and three for $gg$ fusion.  Each diagram includes a color factor $C$ and a scattering amplitude $\mathcal{A}$.  The generic matrix element for each process is \citep{PhysRevD.14.1536}
\begin{eqnarray}
\mathcal{M}_{qq} &=&  C_{qq} \mathcal{A}_{qq} \;, \\
\mathcal{M}_{gg} &=& C_{gg,\hat{s}} \mathcal{A}_{gg,\hat{s}}  + C_{gg,\hat{t}} \mathcal{A}_{gg,\hat{t}} + C_{gg,\hat{u}} \mathcal{A}_{gg,\hat{u}} \; .
\end{eqnarray}

As previously mentioned, there is one diagram only for $q \overline q \rightarrow Q \overline Q$, thus a single amplitude, $\mathcal{A}_{qq}$.  However, there are three diagrams for $gg \rightarrow Q \overline Q$ at leading order, the 
$\hat{s}$, $\hat{t}$ and $\hat{u}$ channels.  In terms of the Dirac spinors $u$ and $v$, the individual amplitudes are
\begin{eqnarray}
\mathcal{A}_{qq} &=& \frac{g_s^2}{\hat{s}} [\overline{u}(p^\prime) \gamma_\mu v(p)][\overline{v}(k) \gamma^\mu u(k^\prime)] \;, \\
\mathcal{A}_{gg,\hat{s}} &=& - \frac{g_s^2}{\hat{s}} \Big\{ -2k^\prime \cdot \epsilon(k) [\overline{u}(p^\prime) \epsilon\!\!\!/(k^\prime) v(p)] \nonumber \\
&+&2 k\cdot \epsilon(k^\prime) [\overline{u}(p^\prime) \epsilon\!\!\!/(k) v(p)] \nonumber \\
&+& \epsilon(k) \cdot\epsilon(k^\prime)[\overline{u}(p^\prime) (k\!\!\!/^\prime - k\!\!\!/) v(p)]\Big\} \;, \\
\mathcal{A}_{gg,\hat{t}} &=& -\frac{g_s^2}{\hat{t}-M^2} \overline{u}(p^\prime) \epsilon\!\!\!/(k^\prime) (k\!\!\!/ -p\!\!\!/ +M) \epsilon\!\!\!/(k) v(p) \;, \\
\mathcal{A}_{gg,\hat{u}} &=& -\frac{g_s^2}{\hat{u}-M^2} \overline{u}(p^\prime) \epsilon\!\!\!/(k) (k\!\!\!/^\prime -p\!\!\!/ +M) \epsilon\!\!\!/(k^\prime) v(p) \;.
\end{eqnarray}
Here $g_s$ is the gauge coupling, $M$ is the mass of heavy quark ($m_c$ for charm and $m_b$ for bottom), $\epsilon$ represents the gluon polarization vectors, $\gamma^\mu$ are the gamma matrices, $k^\prime$ ($k$) is the momentum of initial state light quark (antiquark) or gluon, and $p^\prime$ ($p$) is the momentum of final sate heavy quark (antiquark). 

The amplitudes are separated according to the $J_z$ of the final state, $J_z = 0$ or $J_z = \pm 1$ .  The total amplitudes are calculated for each final state $J_z$ while averaging over the polarization of the initial gluons or the spin of the light quarks, depending on the process, in the spirit of the CEM.  

The squared matrix elements, $|\mathcal{M}|^2$, are calculated for each $J_z$.  The color factors, $C$, are calculated from the SU(3) color algebra and are independent of the polarization \citep{PhysRevD.14.1536}.  They are
\begin{eqnarray}
|C_{qq}|^2 = 2 \;, |C_{gg,\hat{s}}|^2 = 12 \;, \nonumber \\ 
|C_{gg,\hat{t}}|^2 = \frac{16}{3} \;, |C_{gg,\hat{u}}|^2 = \frac{16}{3} \; .
\end{eqnarray}
\begin{eqnarray}
C_{gg,\hat{s}}^*C_{gg,\hat{t}} = +6 \;, C_{gg,\hat{s}}^*C_{gg,\hat{u}} = -6 \;, \nonumber \\
C_{gg,\hat{t}}^*C_{gg,\hat{u}} = -\frac{2}{3} \; .
\end{eqnarray}
The total squared amplitudes for a given $J_z$ state,
\begin{eqnarray}
|\mathcal{M}_{qq}^{J_z}|^2 &=&  |C_{qq}|^2 |\mathcal{A}_{qq}|^2 \;, \\
|\mathcal{M}_{gg}^{J_z}|^2 &=&  |C_{gg,\hat{s}}|^2 |\mathcal{A}_{gg,\hat{s}}|^2 + |C_{gg,\hat{t}}|^2 |\mathcal{A}_{gg,\hat{t}}|^2 \nonumber \\
&+& |C_{gg,\hat{u}}|^2 |\mathcal{A}_{gg,\hat{u}}|^2 + 2  C_{gg,\hat{s}}^*C_{gg,\hat{t}} \mathcal{A}_{gg,\hat{s}}^*\mathcal{A}_{gg,\hat{t}} \nonumber \\
&+& 2 C_{gg,\hat{s}}^*C_{gg,\hat{u}} \mathcal{A}_{gg,\hat{s}}^*\mathcal{A}_{gg,\hat{u}} \nonumber \\ 
&+& 2 C_{gg,\hat{t}}^*C_{gg,\hat{u}} \mathcal{A}_{gg,\hat{t}}^*\mathcal{A}_{gg,\hat{u}} \;,
\end{eqnarray}
are then used to obtain the partonic cross sections by integrating over solid angle:
\begin{eqnarray}
\hat{\sigma}_{ij}^{J_z} = \int d\Omega \Big( \frac{1}{8\pi} \Big)^2 \frac{|\mathcal{M}_{ij}^{J_z}|^2}{\hat{s}} \sqrt{1-\frac{4M^2}{\hat{s}}} \; .
\end{eqnarray}

\begin{figure}
\centering
\includegraphics[width=\columnwidth]{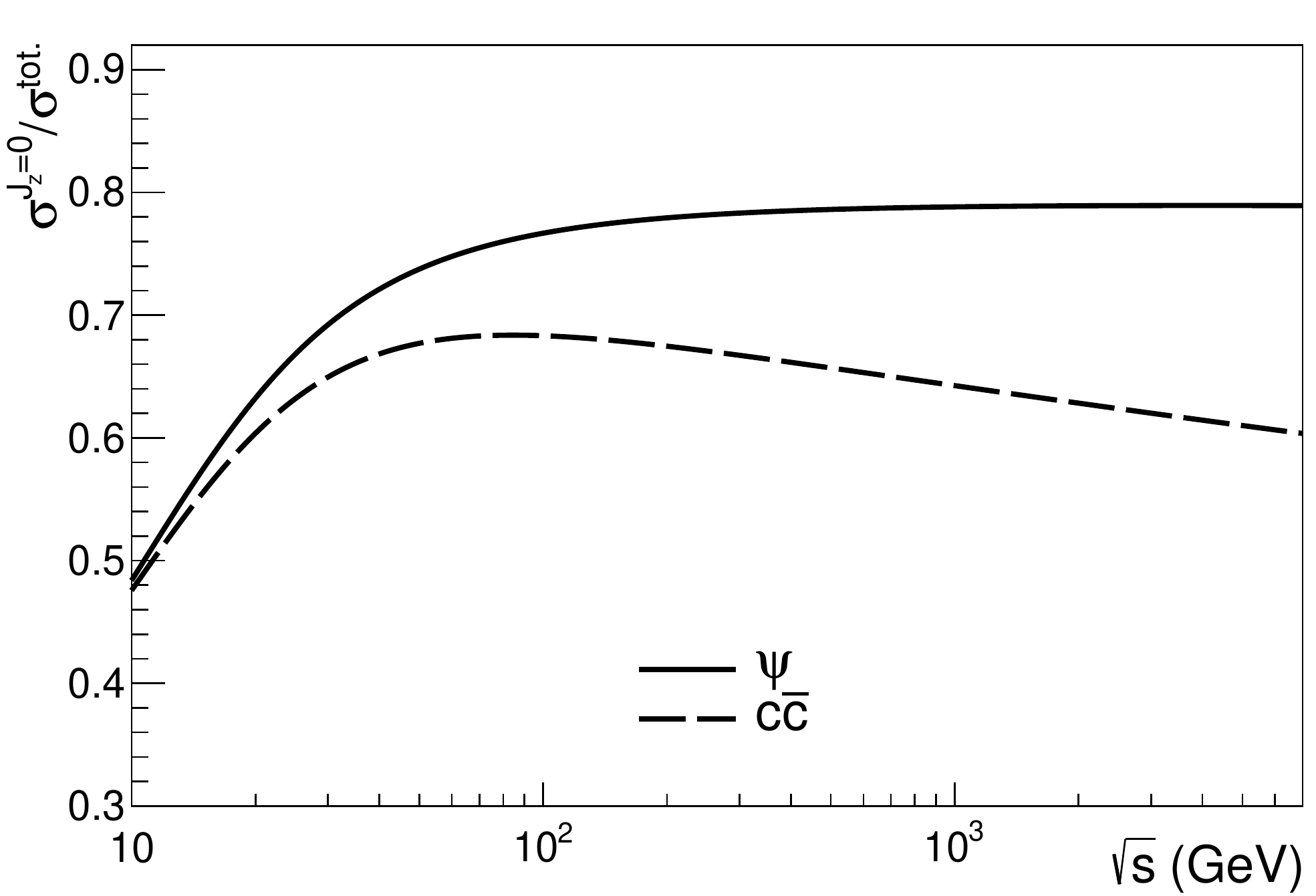}
\caption{\label{LongToUnpolProductionCharm} The energy dependence of the longitudinal fraction for production of charmonium (solid) and $c\overline{c}$ (dashed).}
\end{figure}

The individual partonic cross sections for the longintudinal and transverse polarizations are
\begin{eqnarray}
\hat{\sigma}_{q\overline{q}}^{J_z = 0} (\hat{s}) &=& \frac{16\pi \alpha_s^2}{27 \hat{s}^2} M^2 \chi \;, \\
\hat{\sigma}_{q\overline{q}}^{J_z = \pm 1}  (\hat{s}) &=& \frac{4\pi \alpha_s^2}{27 \hat{s}^2} \hat{s} \chi\;, \\
\hat{\sigma}_{gg}^{J_z = 0} (\hat{s}) &=& \frac{\pi \alpha_s^2}{12\hat{s}} \Big[ \Big(4-\frac{31M^2}{\hat{s}}+\frac{33M^2}{\hat{s}-4M^2} \Big) \chi \\
&+& \Big(\frac{4M^4}{\hat{s}^2} +\frac{31M^2}{2\hat{s}} -\frac{33M^2}{2(\hat{s}-4M^2)} \Big) \ln \frac{1+\chi}{1-\chi} \Big] \;, \nonumber \\
\hat{\sigma}_{gg}^{J_z = \pm 1} (\hat{s}) &=& \frac{\pi \alpha_s^2}{24\hat{s}} \Big[ -11\Big(1+\frac{3M^2}{\hat{s}-4M^2} \Big) \chi \\
&+& \Big( 4+\frac{M^2}{2\hat{s}}+33\frac{M^2}{2(\hat{s}-4M^2)} \Big) \ln \frac{1+\chi}{1-\chi} \Big] \nonumber \;,
\end{eqnarray}
where $\chi = \sqrt{1-4M^2/\hat{s}}$. The sum of these results, $\hat{\sigma}_{ij}^{J_z = 0} + \hat{\sigma}_{ij}^{J_z = +1} + \hat{\sigma}_{ij}^{J_z = -1}$, is equal to the total partonic cross section \citep{Combridge:133166}:
\begin{eqnarray}
\hat{\sigma}_{q\overline{q}}^{\text{tot.}} (\hat{s}) &=& \frac{8\pi \alpha_s^2}{27 \hat{s}^2} (\hat{s} + 2M^2) \chi \;, \\
\hat{\sigma}_{gg}^{\text{tot.}} (\hat{s}) &=& \frac{\pi \alpha_s^2}{3\hat{s}} \Big[ -\Big( 7+\frac{31M^2}{\hat{s}}\Big) \frac{1}{4} \chi \nonumber \\
&&+ \Big( 1+\frac{4M^2}{\hat{s}}+\frac{M^4}{\hat{s}^2} \Big) \ln \frac{1+\chi}{1-\chi} \Big] \;.
\end{eqnarray}
Having computed the polarized $Q\overline{Q}$ production cross section at the parton level, we then convolute the partonic cross sections with the parton distribution functions (PDFs) to obtain the hadron-level cross section $\sigma$ as a function of $\sqrt{s}$ using Eq.~(\ref{cem_sigma}), and the rapidity distribution, $d\sigma/dy$, using Eq.~(\ref{cem_rapdity}). We employ the CTEQ6L1 \citep{Pumplin:2002vw} PDFs in this calculation and the running coupling constant $\alpha_s = g_s^2/(4\pi)$ is calculated at the one-loop level appropriate for the PDFs.  We assume that the polarization is unchanged by the transition from the parton level to the hadron level, consistent with the CEM that the linear momentum is unchanged by hadronization.  This is similar to the assumption made in NRQCD that once a $c \overline c$ is produced in a given spin state, it retains that spin state when it becomes a $J/\psi$.

\begin{figure}
\centering
\includegraphics[width=\columnwidth]{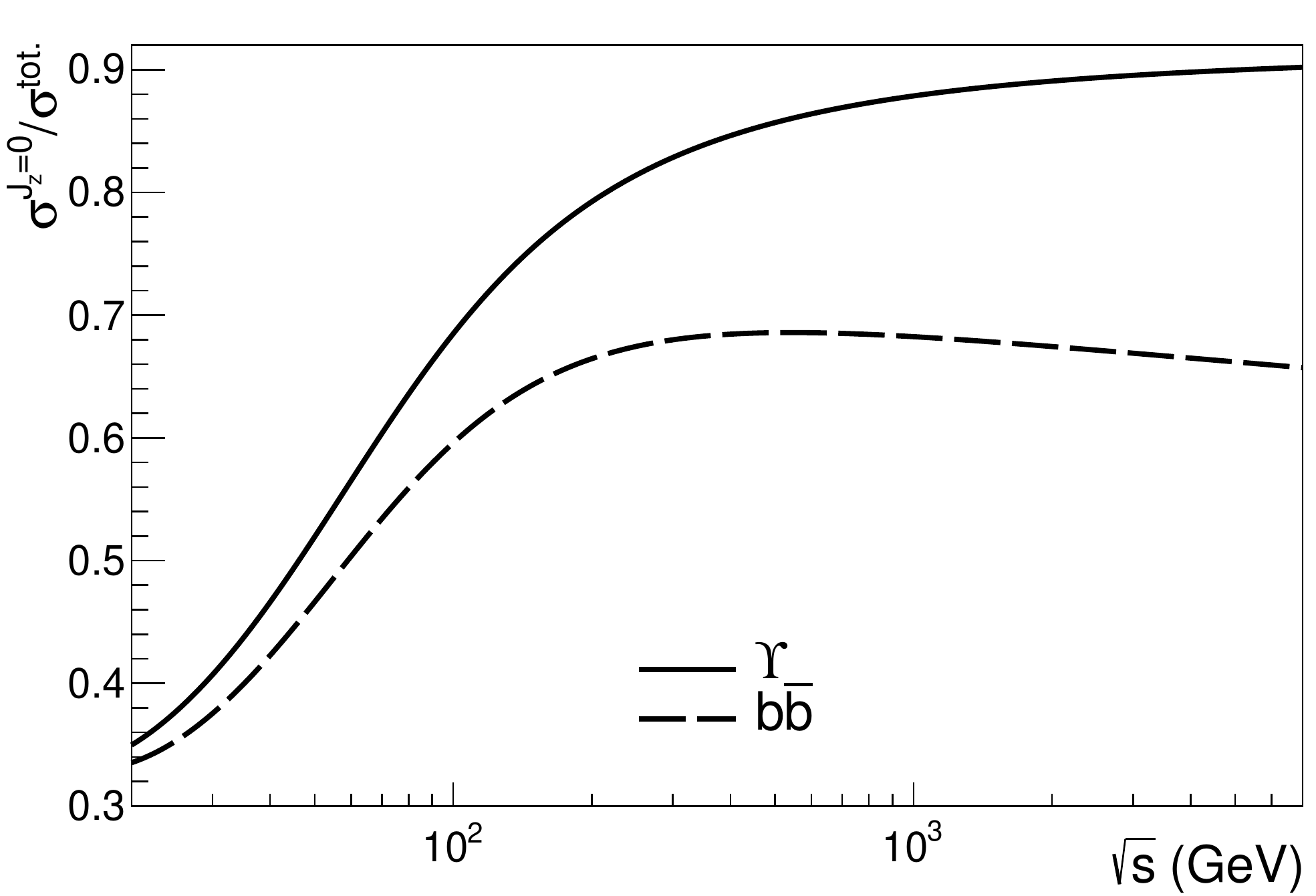}
\caption{\label{LongToUnpolProductionBottom} The energy dependence of the longitudinal fraction for production of bottomonium (solid) and $b\overline{b}$ (dashed).  The result is shown above 20 GeV to be above the $B\overline B$ threshold.}
\end{figure}

\section{Results}

Since this is a LO calculation, we can only calculate the CEM polarization as a function of $\sqrt{s}$ and $y$ but not $p_T$ which will require us to go to NLO. However, the charm rapidity distribution at LO is similar to that at NLO \citep{Vogt:1995zf}. The same is true for $J/\psi$ production in the CEM. The only difference would be a rescaling of the parameter $F_Q$ based on the ratio NLO/LO using the NLO scale determined in Nelson {\it et al.} \citep{NVF}.  The CEM results are in rather good agreement with the data from $p+p$ collisions \citep{NVF}.  

We present the results as ratios of the cross section with $J_z = 0$ to the total cross section. Taking the ratio has the benefit of being independent of $F_Q$. In the remainder of this section, we discuss the energy dependence of the total cross section ratios for both charmonium and bottomonium (in the general sense as being in the mass range below the $H \overline H$ threshold) as well as for $c \overline c$ and $b \overline b$, integrated over all invariant mass.  We show the ratios for charmonium and bottomonium production as a function of rapidity for selected energies.  Finally, we discuss the sensitivity of our results to the choice of proton parton densities.


\subsection{Energy dependence of the longitudinal polarization fraction}

In this section, we compare the energy dependence of the fraction $\sigma^{J_z = 0}/\sigma^{\rm tot.}$ as a function of center of mass energy in $p+p$ collisions in Figs.~\ref{LongToUnpolProductionCharm} and \ref{LongToUnpolProductionBottom}.  In the case of quarkonium, the integration in Eq.~(\ref{cem_sigma}) is from twice the quark mass to twice the mass of the lowest lying open heavy flavor hadron.  For open heavy flavor, the upper limit of the integral is extended to $\sqrt{s}$.


\subsubsection{Charmonium and $c \overline c$}

In Fig.~\ref{LongToUnpolProductionCharm} the charmonium production cross section is calculated by integrating the invariant mass of the $c\overline{c}$ pair from $2m_{c}$ ($m_c=1.27$~GeV) to $2m_{D^0}$ ($m_{D^0}=1.86$~GeV) in Eq.~(\ref{cem_sigma}).  We see that $\psi$ production (solid curve in Fig.~\ref{LongToUnpolProductionCharm}) is more than 50\% longitudinally polarized for $\sqrt{s} > 10$~GeV.  At $\sqrt{s} > 100$~GeV, the production ratio saturates at a longitudinal polarization fraction of 0.80.

The behavior of the total $c\overline{c}$ production fraction (dashed curve in Fig.~\ref{LongToUnpolProductionCharm}) is quite different.  Instead of saturating, like the charmonium ratio, it reaches a peak of 0.68 at $\sqrt{s}=84$ GeV and then begins decreasing. This is because of the approximate helicity conservation at the parton level for $M/\sqrt{\hat{s}} \ll 1$.  The narrow integration range of charmonium production assures that charmonium production never enters this region, keeping charmonium longitudinally polarized.


\subsubsection{Bottomonium and $b \overline{b}$}

\begin{figure}
\centering
\includegraphics[width=\columnwidth]{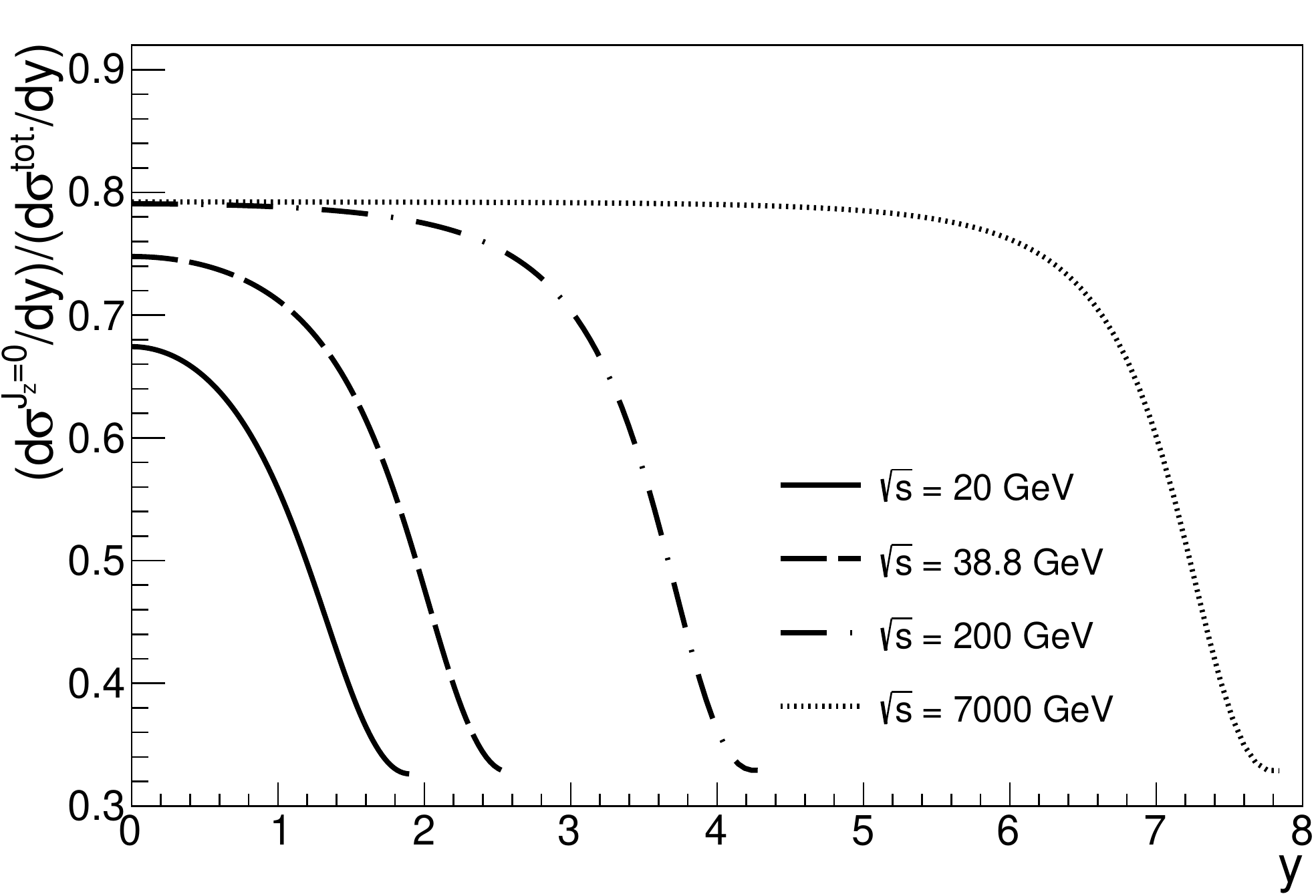}
\caption{\label{LongToUnpolPsi} The rapidity dependence of the longitudinal fraction for production of charmonium at $\sqrt{s}=20$ GeV (solid), 38.8 GeV (dashed), 200 GeV (dot-dashed), and 7000 GeV (dotted). The distributions are symmetric around $y=0$.}
\end{figure}

The results for bottomonium and $b \overline b$ production are shown in Fig.~\ref{LongToUnpolProductionBottom}.  Here, the integral over the pair invariant mass is assumed to be from $2m_{b}$ ($m_b = 4.75$~GeV) to $2m_{B^0}$ ($m_{B^0}=5.28$~GeV).  For the more massive bottom quarks, the pairs start out transversely polarized for $\sqrt{s} < 40$~GeV.  Bottomonium production becomes dominated by longitudinal polarization but the ratio saturates at 0.90 for $\sqrt{s}$ of $\sim 1$~TeV, higher than the charmonium ratio at the same energy.  The smaller longitudinal fraction at lower $\sqrt{s}$ for bottomonium is because of $q \overline q$ dominance of the total cross section at these energies.  As the $gg$ contribution rises, the longitudinal fraction increases.

We note that the point at which the bottomonium fraction is $\sim 0.50$, $\sqrt{s} = 46.3$~GeV, is similar to the lowest energy at which $\Upsilon$ polarization has been measured, $\sqrt{s_{NN}} = 38.8$ GeV.  The E866/NuSea Collaboration measured the polarization of bottomonium production in $p+$Cu and found no polarization at low $p_T$ in the Collins-Soper frame \citep{PhysRevLett.86.2529}. This result is compatible with our own because at leading order, the polarization axes in the helicity frame, the Collins-Soper frame, and the Gottfried-Jackson frame frame are coincident \citep{Faccioli:2010kd}.

Likewise, the turnover in the $c \overline c$ polarization is also observed for $b \overline b$ but at a much higher energy, $\sqrt{s}=550$~GeV.  Although the energy scale is higher, the peak in the $b \overline b$ polarization ratio is almost the same as that for $c \overline c$, 0.69. 


\begin{figure}
\centering
\includegraphics[width=\columnwidth]{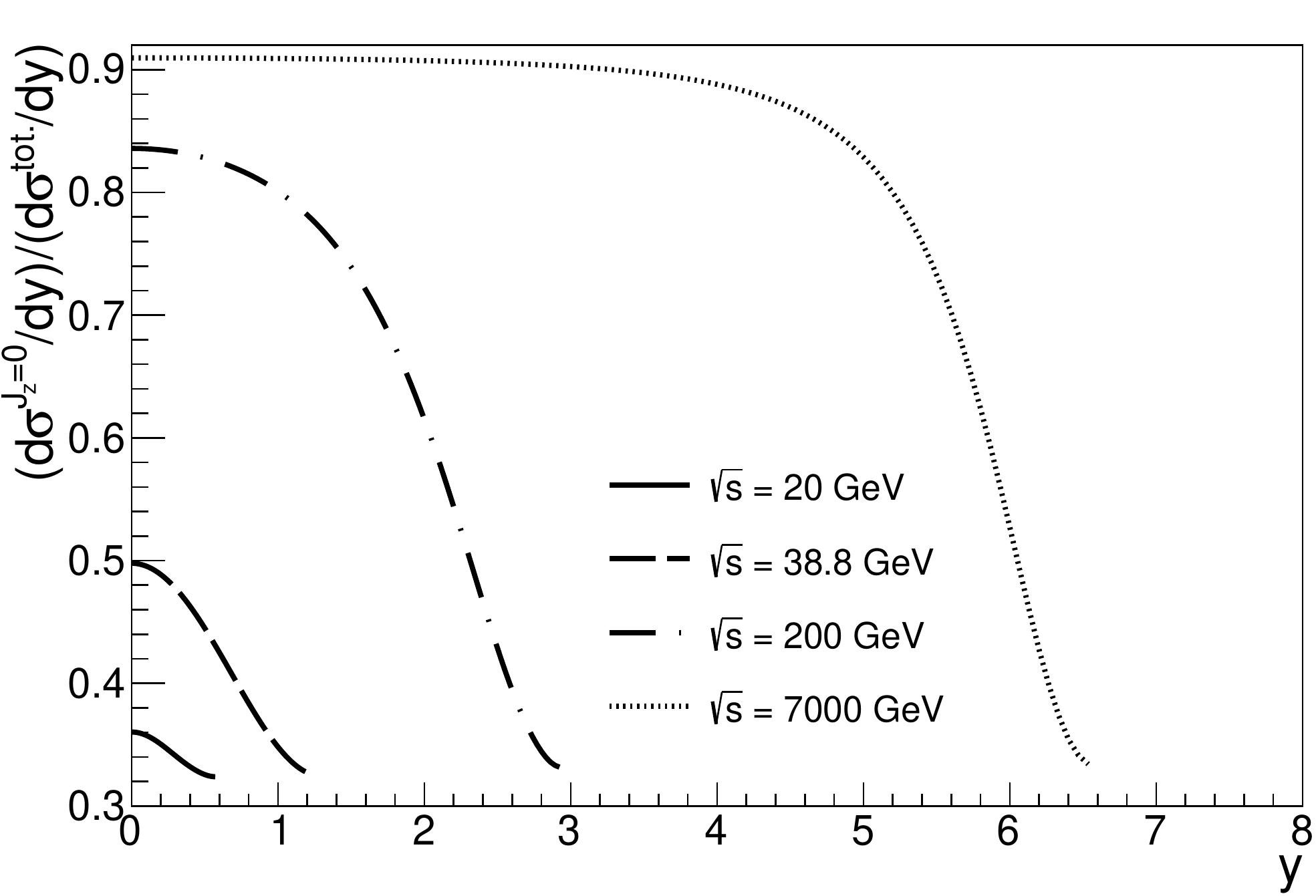}
\caption{\label{LongToUnpolUpsilon} The rapidity dependence of the longitudinal fraction for production of bottomonium at $\sqrt{s}=20$ GeV (solid), $\sqrt{s}=38.8$ GeV (dashed), 200 GeV (dot-dashed), and 7000 GeV (dotted). The distributions are symmetric around $y=0$.}
\end{figure}

\subsection{Rapidity dependence of the longitudinal polarization fraction}

We now turn to the rapidity dependence of our result, shown in Figs.~\ref{LongToUnpolPsi} and \ref{LongToUnpolUpsilon}.  Four representative energies are chosen to illustrate.  The lowest values, $\sqrt{s} = 20$ and 38.8~GeV were the highest available fixed-target energies at the CERN SPS for ion beams and the FNAL Tevatron for proton beams.  The higher energies, $\sqrt{s} = 0.2$ and 7~TeV are energies available at the BNL RHIC and CERN LHC facilities.  The results are presented for positive rapidity only because the rapidity distributions are symmetric around $y=0$ in $p+p$ collisions.

\subsubsection{Charmonium}

The rapidity dependence for the charmonium longitudinal polarization fraction is shown in Fig.~\ref{LongToUnpolPsi}. The results are given up to the kinematic limits of production.  The longitudinal fraction is greatest at $y = 0$ and decreases as $|y|$ increases. For the highest energies, where the longitudinal polarization has saturated in Fig.~\ref{LongToUnpolProductionCharm}, the ratio is flat over a wide range of rapidity.  The ratio remains greater than 0.50 as long as the $gg$ contribution, with a significant $J_z = 0$ polarization,  dominates production.   As the phase space for charmonium production is approached, the $q \overline q$ channel, predominantly transversely polarized, begins to dominate, causing the ratio to drop to a minimum of $\sim 0.30$.


\subsubsection{Bottomonium}

The behavior of the bottomonium ratio as a function of rapidity, shown in Fig.~\ref{LongToUnpolUpsilon}, is similar to that of charmonium.  The higher mass scale, however, reduces the kinematic range of the calculation.  It also results in near transverse $(J_z = \pm 1)$ polarization of bottomonium at fixed-target energies.  The calculation at $\sqrt{s} = 38.8$~GeV shows that, at $y=0$, the bottomonium ratio is consistent with no polarization, as measured by E866/NuSea \cite{PhysRevLett.86.2529}.  At $\sqrt{s} = 20$~GeV, not far from production threshold, bottomonium is transversely polarized in the narrow rapidity range of production. 


\subsection{Sensitivity to the proton PDFs}
\label{PDF}

We have tested the sensitivity of our results to the choice of PDFs used in the calculation.  Since not many new LO proton PDFs are currently being made available, we compare our CTEQ6L1 results with calculations using the older GRV98 LO \citep{Gluck:1998xa} set.  We can expect the ratio to be the most sensitive to the choice of proton PDF because the PDFs can change the balance of $gg$ to $q \overline q$ production, especially at lower $\sqrt{s}$ where the $x$ values probed by the calculations are large, $x \sim 0.1$.  In particular, bottomonium production at $\sqrt{s} = 20$~GeV is most likely to be sensitive to the choice of PDF since the $q \overline q$ contribution is large at this energy.  The results should, on the other hand, be relatively insensitive to the chosen mass and scale values since these do not strongly affect the relative contributions of $gg$ and $q \overline q$.

This is indeed the case, for bottomonium production at $\sqrt{s} = 20$~GeV, close to the production threshold, the largest difference in the longitudinal ratio for the two PDF sets is 36\% at $y=0$.  The sensitivity arises because the $gg$ contribution is predominantly produced with $J_z = 0$ while the $q \overline q$ contribution is primarily produced with $J_z = \pm 1$.  By $\sqrt{s} = 38.8$~GeV, the difference in the results is reduced to 20\%.  At collider energies, the difference is negligible.  Since the $gg$ contribution is dominant for charmonium already at $\sqrt{s} = 20$~GeV, the charmonium production ratio is essentially independent of the choice of proton PDF.  Thus, away from production threshold, the results are robust with respect to the choice of PDF.


\section{Conclusion}

We have presented the energy and rapidity dependence of the polarization of heavy quarkonium production in $p+p$ collisions in the Color Evaporation Model.  We find the quarkonium polarization to be longitudinal at most energies and around central rapidity while the polarization becomes transverse as the kinematic limits of the calculation, where $q \overline q$ production is dominant, are approached.

We note that the partonic cross sections, sorted by $J_z$ in this calculation, are still mixtures of total angular momentum $J$ and orbital angular momentum $L$ states. So there is no immediate connection between these ratios and the lambda parameter of the data. In future work, we will extract the $S=1$, $L=0$ contribution from the partonic cross sections to narrow down into three distinct angular momentum states of $J=1$ in order to give predictions for the polarization parameter $\lambda_\theta$ \citep{Faccioli:2010kd}. 

Because we have performed a leading order calculation, we cannot yet speak to the $p_T$ dependence of the quarkonium polarization. We will address the $p_T$ dependence in a separate publication.


\section{Acknowledgemets}
We thank F. Yuan for valuable discussions throughout this work. This work was performed under the auspices of the U.S. Department of Energy by Lawrence Livermore National Laboratory under Contract DE-AC52-07NA27344 and supported by the U.S. Department of Energy, Office of Science, Office of Nuclear Physics (Nuclear Theory) under contract number DE-SC-0004014.

\bibliography{CEM.bib}

\end{document}